\documentclass[apj]{emulateapj} \setkeys{Gin}{width=0.48\textwidth}
\usepackage{graphicx}

\newcommand{\hmpc}{\ifmmode{h^{-1}\,\hbox{Mpc}}\else{$h^{-1}$\thinspace Mpc}\fi}

\newcommand{\kms}{\ifmmode{\,\hbox{km\,s}^{-1}}\else {\rm\,km\,s$^{-1}$}\fi}
\newcommand{\msun}{{\rm\,M_\odot}}

\begin{document}
\title{The Density Structure of Simulated Stellar Streams}
\shorttitle{Stellar Streams and Sub-halos}
\shortauthors{Carlberg}
\author{Raymond G. Carlberg}
\affil{Department of Astronomy \& Astrophysics, University of Toronto, Toronto, ON M5S 3H4, Canada} 
\email{carlberg@astro.utoronto.ca }

\begin{abstract}
Star particles in a set of dense clusters are self-consistently evolved
within an LCDM dark matter distribution with an n-body code.
The clusters are  started on nearly circular orbits in the more massive sub-halos. 
Each cluster develops a stellar tidal stream, initially within its original sub-halo.
When a sub-halo merges into the main halo 
the early time stream is dispersed as a somewhat chaotic thick stream, 
roughly the width of the orbit of the cluster in the sub-halo.
Once the cluster  orbits freely in the main halo the star stream 
forms a thin stream again, usually resulting in a thin stream surrounded by a wider distribution of 
star particles lost at earlier times.
To examine the role of the lower mass dark matter sub-halos 
in the creation of density variations along
the thin tidal star streams two realizations of the simulation are run 
with and without a normal cold dark matter sub-halo population below $4\times 10^8\msun$ 
About 70(40)\% of thin streams show density variations that are 2(5) times 
the star count noise level, irrespective of the presence or absence of low mass sub-halos. 
A counts-in-cells analysis (related to the two-point correlation function and power spectrum)
of the density along nearly 8000\degr\ of streams in the two well matched models
 finds that the full sub-halo population leads
to slightly larger, but statistically significant, density fluctuations on scales of 2-6\degr.
\end{abstract}


\section{INTRODUCTION}
\nobreak

The galactic halo globular clusters observable today likely formed along with other star clusters in a rotating disk 
of gas in many separate dark matter sub-halos, the high redshift forms of dwarf galaxies, that later accreted onto the growing Milky Way halo
\citep{Grudic:18,Bekki:19,Li:19,ReinaCampos:19}.
As the residual gas of formation was driven away loosely bound clusters were dispersed. The tidal 
fields of of surrounding molecular clouds,  while present, and the galactic tides of the  sub-halo of formation  continued
to heat and disperse stars from the clusters \citep{Chandar:17}.  
Only the densest star clusters survived for more than a few hundred million years.
As they orbited in their initial dark matter halo they began to produce
coherent tidal streams of stars which continued as those halos merged. As a result
the tidal streams are a rich source of information for the current and past gravitational potential of the galaxy.

Analysis of the small scale  density structure of tidal streams  can provide information on 
the numerous low mass dark matter sub-halos that LCDM cosmology 
predicts should be present within our galactic halo \citep{Klypin:99,Moore:99}.
\citet{Ibata:02} and \citet{Johnston:02} showed that individual sub-halo encounters with idealized streams  would
visibly perturb the density along a tidal stream.
The cumulative effects of the expected population of sub-halos 
was explored with simulations and dynamical analysis, first with a simplified uniform stream on a circular orbit 
\citep{YJH:11,Carlberg:13} and later with more realistic orbits and streams \citep{NC:14,EB:15,HK:16,EBD:16}.
An uncertainty in predictions is that disk and bulge of the visible galaxy causes the halo to become more spherical \citep{Dai:18}
and depletes 
the numbers of sub-halos in the inner part of the halo 
\citep{DOnghia:10,FIREdisk:17, KBG:19}
where streams are most readily found \citep{GC:16}.
Ideally the properties of the density variations along the stream
can be used to predict the location of the perturbing sub-halo  \citep{Bonaca:19} or identify
a visible 
component of the galaxy, such as the bar \citep{Pearson:17}, LMC \citep{ErkalOrphan:19} or the
possibility of chaotic orbits in the halo \citep{PriceWhelan:16} as the source of stream perturbations.
The underlying effort is to develop reliable dynamical  tests for the nature
of dark matter through the numbers of sub-halos below $10^8-10^9\msun$ \citep{Viel:05,AHA:13}
which are directly related to the number and scale of density variations along a stream \citep{Bovy:17,BBB:18}.
A set of streams are reliable indicators of the total enclosed mass even when dynamically formed in a cosmological setting
although a single stream can be quite biased \citep{Bonaca:14,Sanderson:17}.

Analytic approaches to characterizing the density variations and gaps
rely on the stream being produced in a halo potential that can
be adequately approximated as static.
In a stationary potential the tidal mass loss variations \citep{Kupper:08,Kupper:12} are 
phase mixed away in a few kiloparsecs of stream length, so density variation analysis needs to discard the near-cluster
part of the stream. Although the last 10 Gyr of the Milky Way's mass assembly
has been sufficiently quiet to allow the build-up of a disk \citep{TO:92,HHC:08} globular clusters have been 
losing stars to tidal streams for essentially a Hubble time and clusters are accreted over
a wide range of times \citep{SagGC:03,SausageGC:19,GaiaEncGC:19} meaning that almost all clusters
have experienced complex gravitational potential histories.
Improving the comparison with observations requires realistically created streams  which in turn requires
numerical simulations of tidal mass loss from globular clusters in an evolving galaxy \citep{NBC:15,SKJ:17,Carlberg:18}. 

This paper presents n-body simulations of the evolution of globular star clusters
within a dark matter halo started from its high redshift state.  The density profiles along the length 
and transverse to the center line of a representative set of streams are measured to help guide
the analysis of streams.
A simulation is run with and without sub-halos below $4\times 10^8\msun$ to examine 
the role of lower mass sub-halos in creating smaller scale density variations in the streams.

\section{Simulation Setup}

The simulation procedure here is based on the methods of  \citet{Carlberg:18} with 
a number of enhancements. 
The starting conditions use the VL2 catalog of dark matter halos found within a larger
scale cosmological simulation that formed a Milky Way-like galaxy \citep{VL2}. The halo catalog, which gives
a characteristic radius, circular velocity, position and velocity for 20048 sub-halos, is used 
to reconstitute the halos as  particle distributions of equilibrium \citet{Hernquist:90} spheres. 
The starting conditions are set up 
at redshift 3.24 and 4.56, ages of 2.08 and 1.39 Gyr. respectively. The standard model recreates all halos with at least
one dark matter particle, which generates 19596 and 19546 of the 20048 halos in the VL2 catalog above this mass limit 
at redshifts 3.2 and 4.6, respectively. 

\subsection{Star cluster setup}

The initial masses of the star clusters are drawn from a power law mass distribution 
$dN/dm \propto M^{-1.5}$, with an upper
mass limit of $2\times 10^6\msun$ and a lower mass limit of $5\times 10^4\msun$. 
Lower mass clusters are not useful to include because internal relaxation causes them to evaporate
fairly quickly, often in their initial dark matter sub-halo, simply leading to a dispersed set of halo stars. The mass function
slope of $-1.5$  is somewhat
shallower than the $-2$ that might be preferred \citep{PZM:00}, but 
ensures that the most massive clusters always  dominate the total mass in the cluster system \citep{FZ:01}. 
The statistical results can be rescaled to a mass function slope of $-2$ if desired. The upper mass limit is inserted simply
as a numerical convenience to limit the number of high mass clusters.
Star clusters are drawn from the mass distribution until  the 
ratio of the total stellar mass to dark matter mass, $\eta$ in the entire simulation rises to $\eta = 3\times 10^{-4}$.
This is the initial $\eta$ value which will be reduced as the clusters lose mass.
\citet{HHH:14} find a current epoch ratio $\eta \simeq 4\times 10^{-5}$ to which the results related to cluster numbers can be scaled if
desired. The redshift 4.6 start begins with $\eta = 1\times 10^{-4}$.

The setup at redshift 3 creates 1433 star clusters distributed over 394 sub-halos above
a mass of $4\times 10^4\msun$. The star particles have a mass of $10\msun$.
This setup will
be used to examine the role of lower mass sub-halos in creating density variations in streams in the last  section of the paper.
The redshift 4.6 start has 710 star clusters distributed over 455 sub-halos with star particles of $5\msun$.
This configuration will be used
to examine the typical mean density structure of streams along and transverse to streams in the next section of the paper.

The clusters are given random radial locations drawn from an exponential disk with a cut-off at 3 scale radii.
The disk scale radius  is set at 20\% of the halo radius of the peak of the circular velocity curve
of the  dark matter sub-halo.  The disks are randomly oriented within each halo.
The larger sub-halos in the redshift 3.2 start contain a few dozen clusters. 
The clusters are started at the local circular velocity
with an additional 3 \kms\ of random velocity added to each direction to limit disk instabilities. 

The star particles are distributed within each cluster using a \citet{King:66} model distribution. 
The King model has a normalized central potential depth of $W_0=7$, which gives a core radius 22 
times smaller than the outer radius of the cluster.
For most of the clusters such a small core is not resolved with the 2 pc gravitational softening used, but
there is no need here to accurately resolve the core.
Individual clusters have a mass assigned from the mass distribution. 
The local tidal radius in the halo is calculated from the classic Jacobi radius, and then reduced by a factor of 3 
so that all clusters initially under-fill their tidal radius. Tidal and internal  heating of the clusters causes them to expand 
to an equilibrium radius appropriate to the local tidal fields.

\subsection{Star Cluster Internal Dynamics}

The star particles are given a gravitational softening of 2 parsecs. The softening
diminishes interactions between individual cluster stars to essentially zero.
A Monte Carlo heating model restores
the interactions to the level that reproduces cluster mass loss and half mass evolution with a Monte Carlo heating.
The heating model is based on the 
Spitzer half-mass relaxation rate \citep{Spitzer:87,BT:08} to allow for different cluster sizes and masses, 
or equivalently virial radius and velocity dispersion. 
The Spitzer relaxation rate is used to calculate the velocity width of
a Gaussian distribution from which random velocities are drawn and added
to each star within the cluster. The heating is normally applied every 5 Myr so that the stars 
have an opportunity to orbit significantly between interactions.
There is no internal source of energy from a shrinking core or hard binaries, so on this model the added velocities 
are an external heat source.

The rate of Monte Carlo heating is calibrated with NBODY6 runs  \citep{Aarseth:99} in a static potential
which then allows the simple heating model to accurately reproduce the mass loss  from the star clusters in the $10^{4-5} \msun$ range.
Ongoing calibration work shows that the heating parameters used in \citet{Carlberg:18} need to be reduced for  clusters
above $10^{5} \msun$.  The redshift 4.6 model is run with a more accurate heating rate for high mass clusters 
(Peter Berczik, Jongsuk Hong, Yohai Meiron, Jeremy Webb private communication).
 Individual particles in the models have a mass of $5-10\msun$, but the heating model is scaled to 
stars of mean mass of around $0.5\msun$, as is appropriate to old clusters. Tests of these clusters orbiting 
in a static potential demonstrate that the results are independent of the star particle masses below about 
$50\msun$ for $4\times 10^5\msun$ clusters.

\begin{figure*}
%\begin{interactive}{animation}{movie60_461.mp4}
\includegraphics[angle=0,scale=2.1]{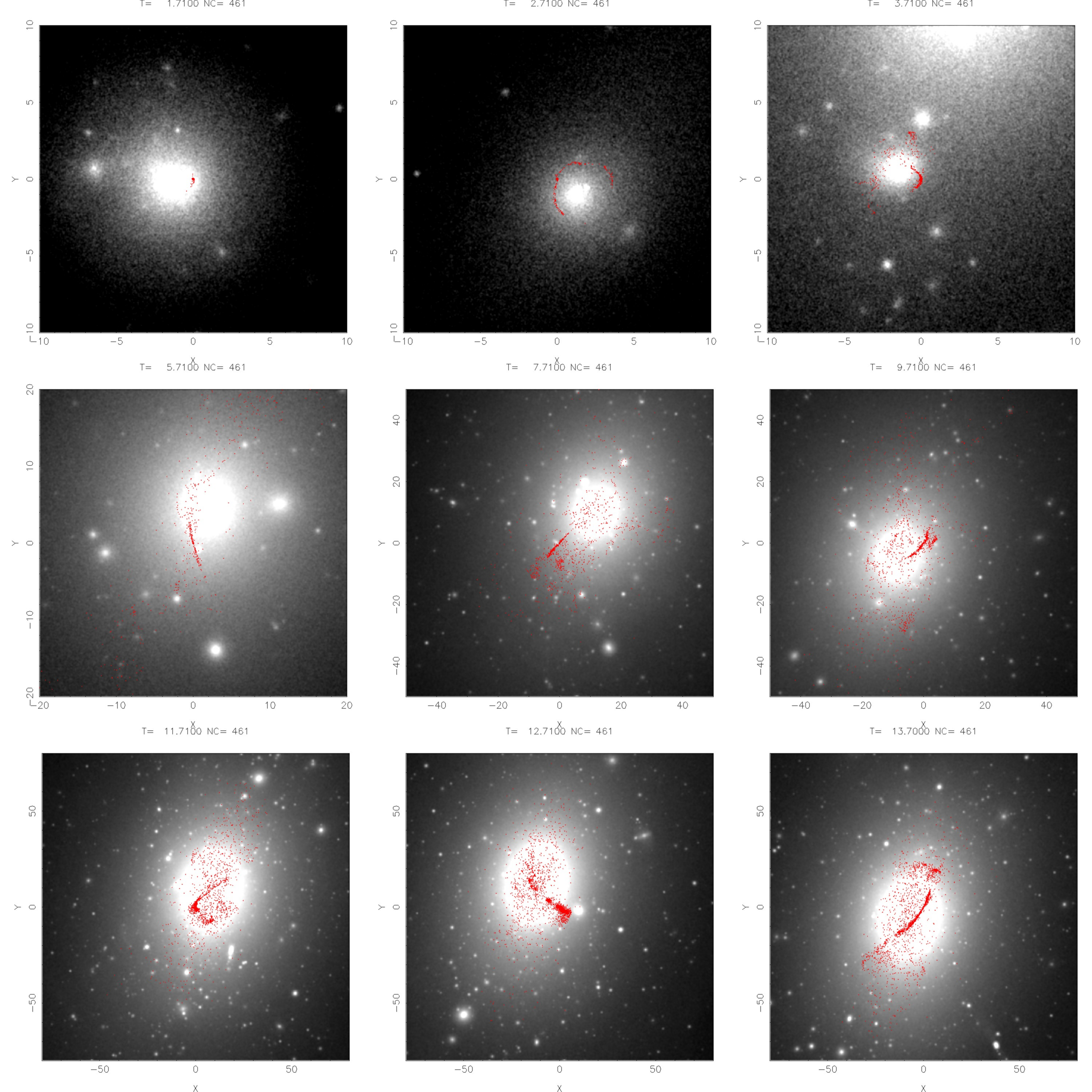}%{sdm499.pdf}
%\includegraphics[angle=0,scale=1.2]{Figure_1.pdf}%{sdm499.pdf}
%\end{interactive}
\caption{The evolution of the tidal stream from a single cluster with time. 
The simulation age in computational units of 1.022 Gyr is indicated above each panel, increasing from the upper left
to the lower right.  Note that the sizes of the plotted boxes increase with time in the set of stills. The
grey scale is proportional to the square root of the projected dark matter density. The image is always centred on the current location 
of the star cluster center. 
The red dots are individual stars from a single cluster.
The movie shows the time evolution.
}
\label{fig_sdm}
\end{figure*}

\begin{figure*}
\begin{center}
\includegraphics[angle=-90,scale=0.8,trim=180 70 180 70,clip=true]{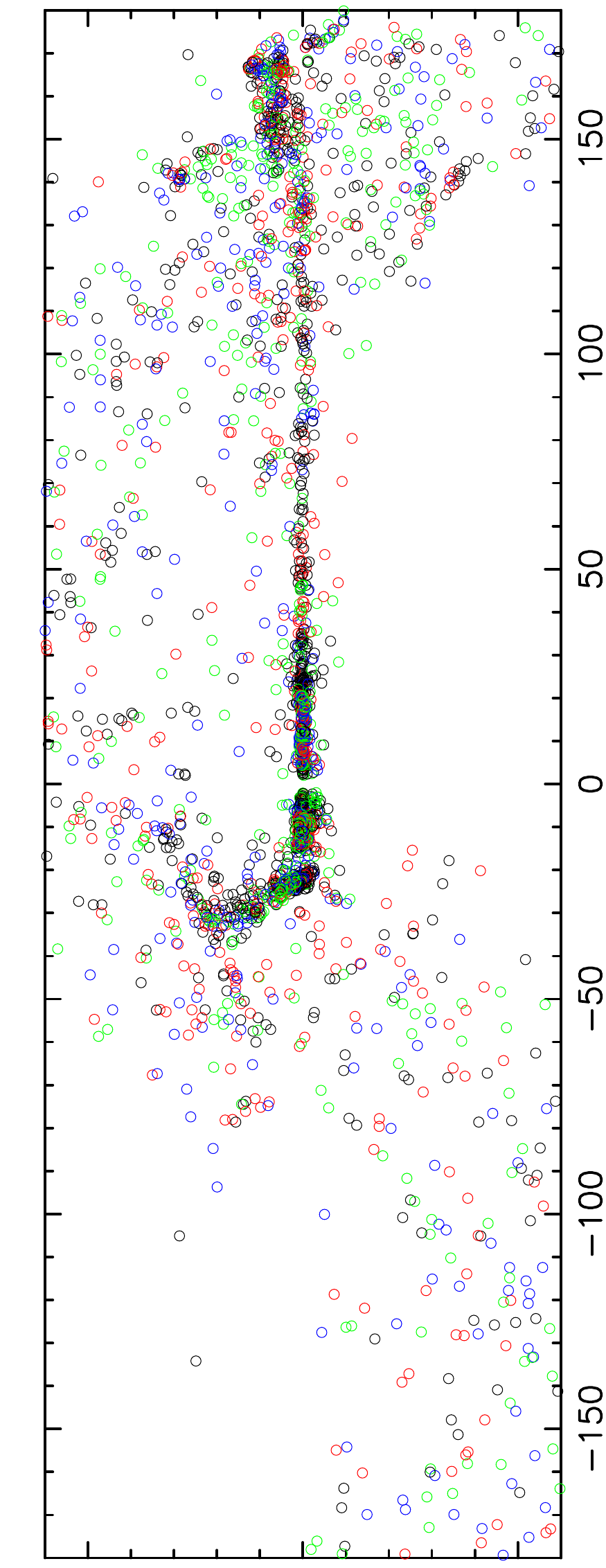}%{aLT1232_9.pdf}}
\put(-510,-185){\includegraphics[angle=-90,scale=0.8,trim=180 70 180 70,clip=true]{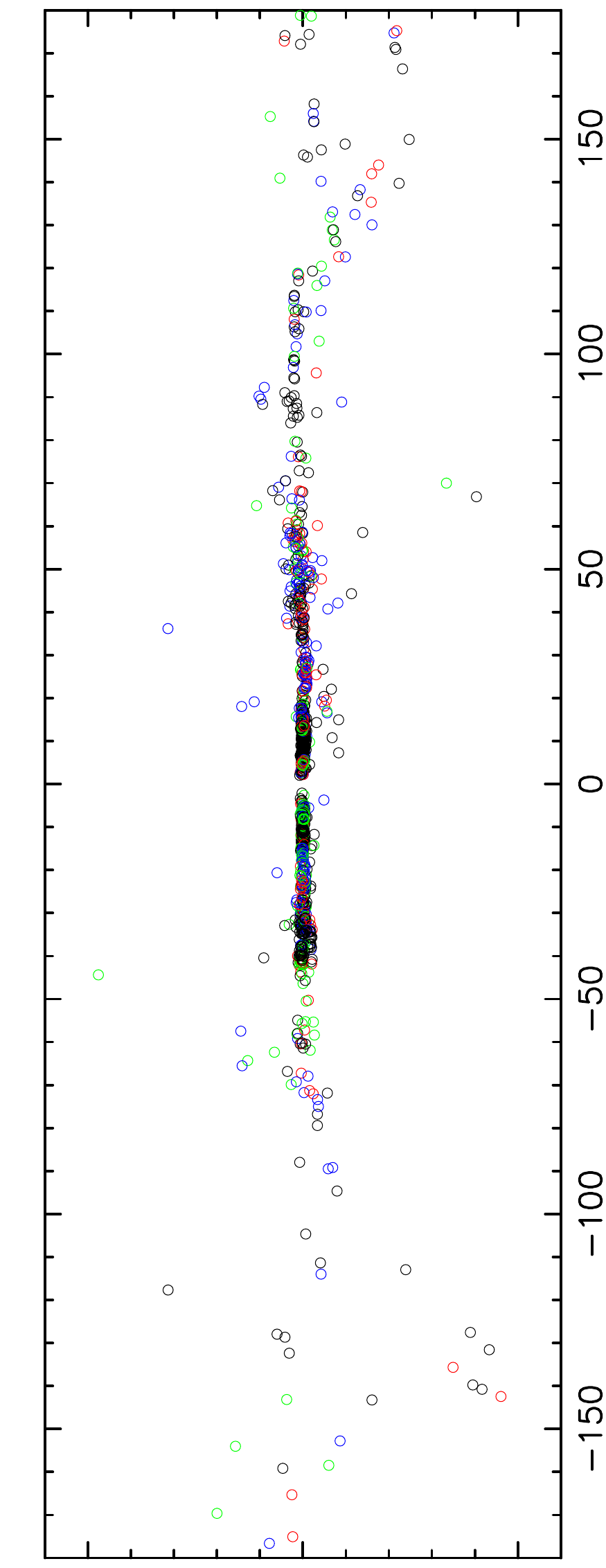}}%{aLT1232_15.pdf}}
%\includegraphics[angle=-90,scale=0.45,trim=180 70 180 70,clip=true]{Figure_2a.pdf}%{aLT1232_9.pdf}}
%\put(-465,-175){\includegraphics[angle=-90,scale=0.45,trim=180 70 180 70,clip=true]{Figure_2b.pdf}}%{aLT1232_15.pdf}}
\end{center}
\caption{The top panel stars from the cluster shown in Figure~\ref{fig_sdm} projected onto the sky with a cylindrical 
projection. The units are degrees with a square grid. The orbital plane is defined by the velocity of the cluster. The stars
are color coded with time time since they were last inside 100 pc of the cluster center, 
0-3.01 Gyr (black), 3.01-6.02 Gyr (red), 6.02-9.03 (green) and 9.03-12.04 (blue) Gyr.
The simulation began at 12.04 Gyr. Stars within 2\degr\ of the cluster center are not plotted.
The bottom panel shows a cluster in which the thin stream is much more dominant.
}
\label{fig_timeout}
\end{figure*}

\begin{figure}
\begin{center}
\includegraphics[angle=-90,scale=0.7,trim=90 30 40 30,clip=true]{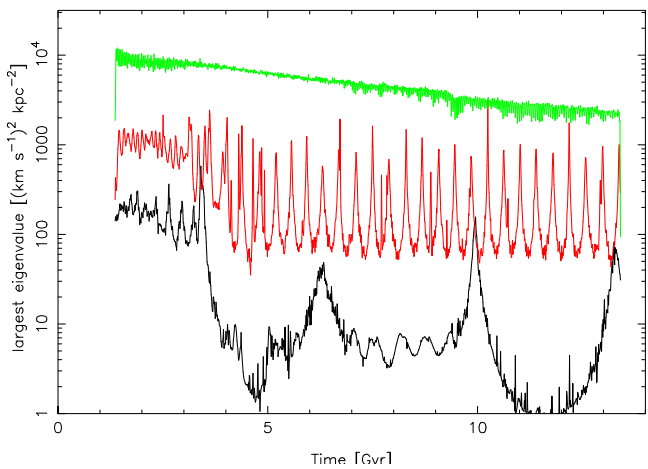}%{tides.pdf}
\end{center}
\caption{The largest eigenvalue of the tidal field matrix as function of time
for three representative clusters, with the cluster shown in Figures~\ref{fig_sdm} and \ref{fig_timeout} plotted in red.
The cluster in the lower panel of Figure~\ref{fig_timeout} is shown in green.
}
\label{fig_tides}
\end{figure}

\begin{figure*}
\begin{center}
\includegraphics[angle=-90,scale=1.3,trim=80 70 120 30,clip=true]{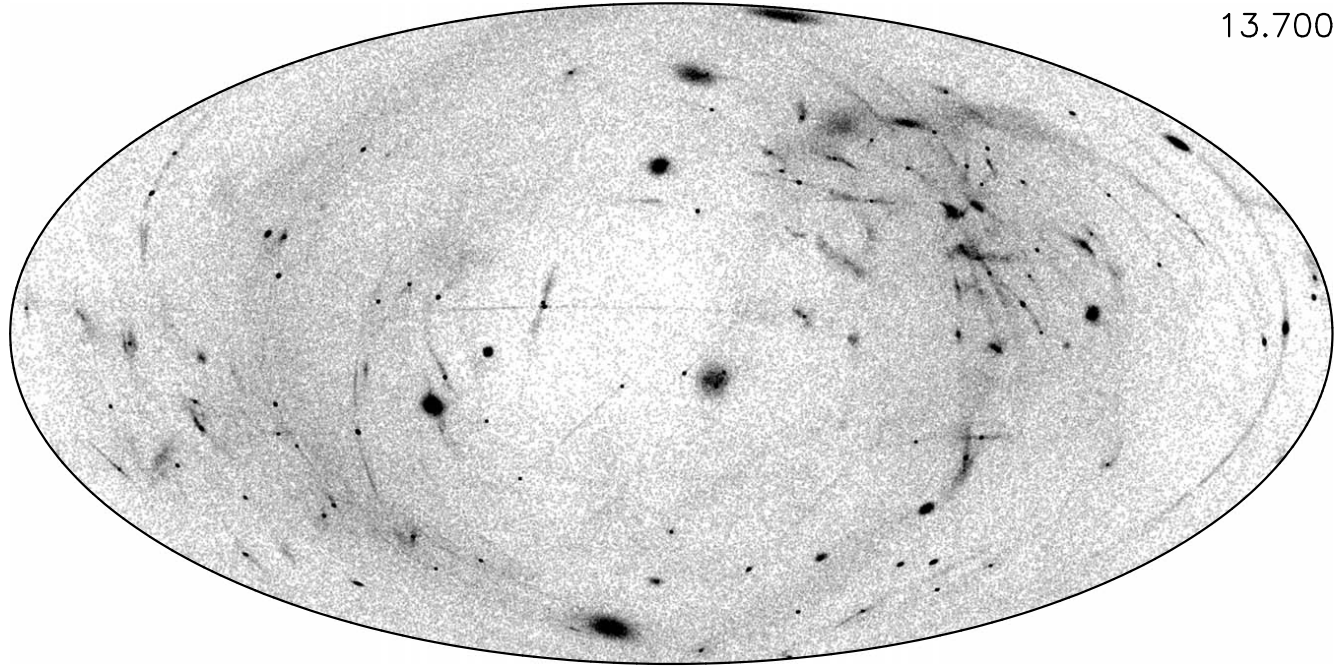}%{ah1232ff.pdf}
\end{center}
\caption{The Hammer-Aitoff projection on to the sky of all the star particles with parent clusters located between 8 and 40 kpc 
of the galactic center at the end of the simulation. The radial range excludes streams that the galactic disk would certainly
interact with and those
that would be too distant to be readily found in current surveys, although expanding the range to 0 to 100 kpc does not substantially alter the plot.}
\label{fig_ah}
\end{figure*}

\begin{figure*}
\begin{center}
\includegraphics[angle=0,scale=2.1]{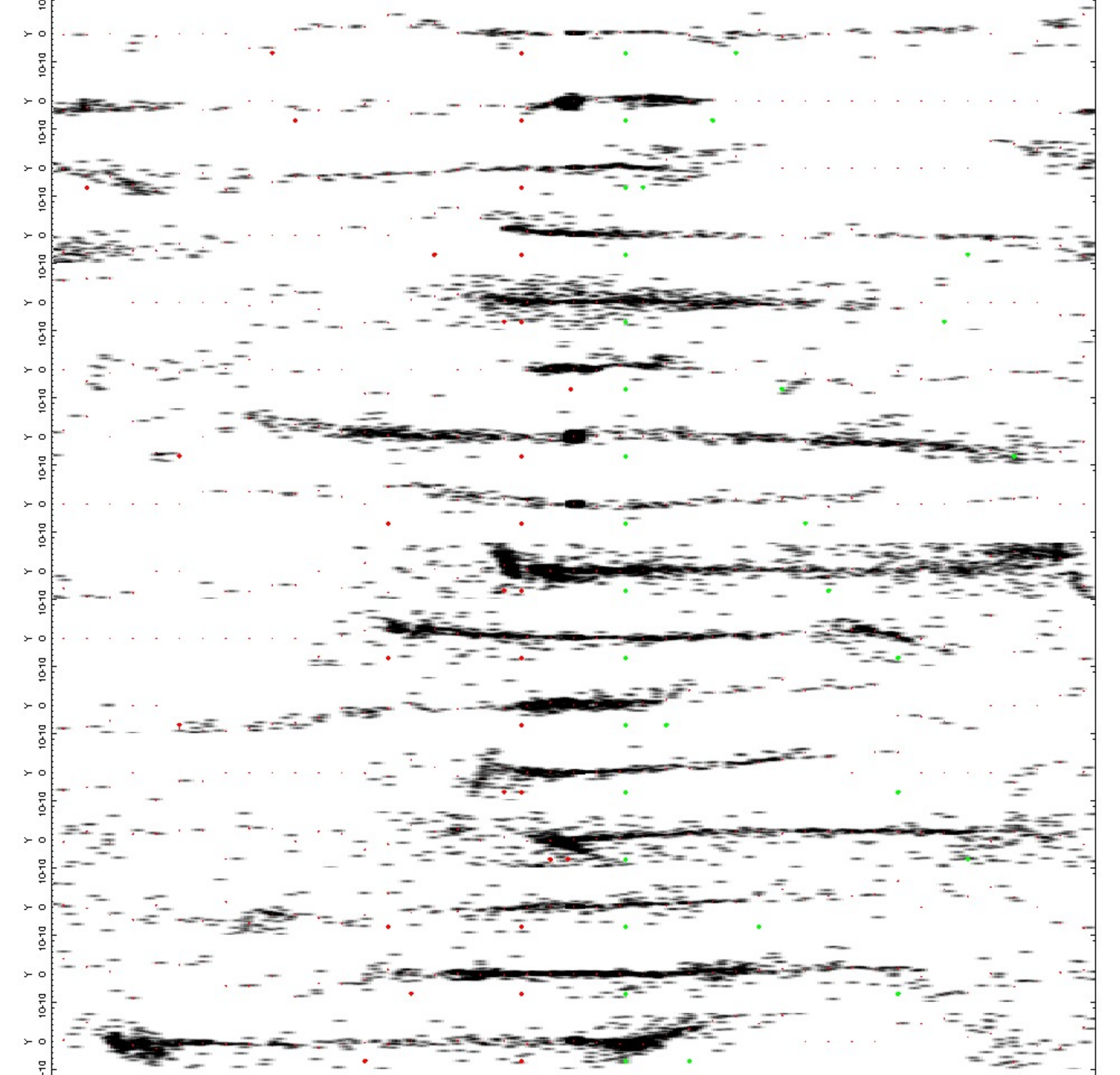}%{montage358.pdf}
\end{center}
\caption{A Mercator projection of all stars from the 16 clusters within 60 kpc of the galactic center with thin streams having
an RMS full width of less than 1.2\degr.
Each stream is rotated into a frame with the equator aligned with the velocity vector of the progenitor star cluster.
The gray scale emphasizes low density regions with a gray scale proportional to the square root
of the number of  particles per $2\times 0.3\degr$ pixel. The scale saturates at a value of 1.5, or 2.25 particles per pixel.
The cluster in Figures~\ref{fig_sdm} and \ref{fig_timeout} and the red line
in Figure~\ref{fig_tides} is the ninth from the top. The black line of Figure~\ref{fig_tides} is the top stream
and the green line is the second from the bottom.
}
\label{fig_in60}
\end{figure*}

\begin{figure*}
\begin{center}
\includegraphics[angle=0,scale=2.1]{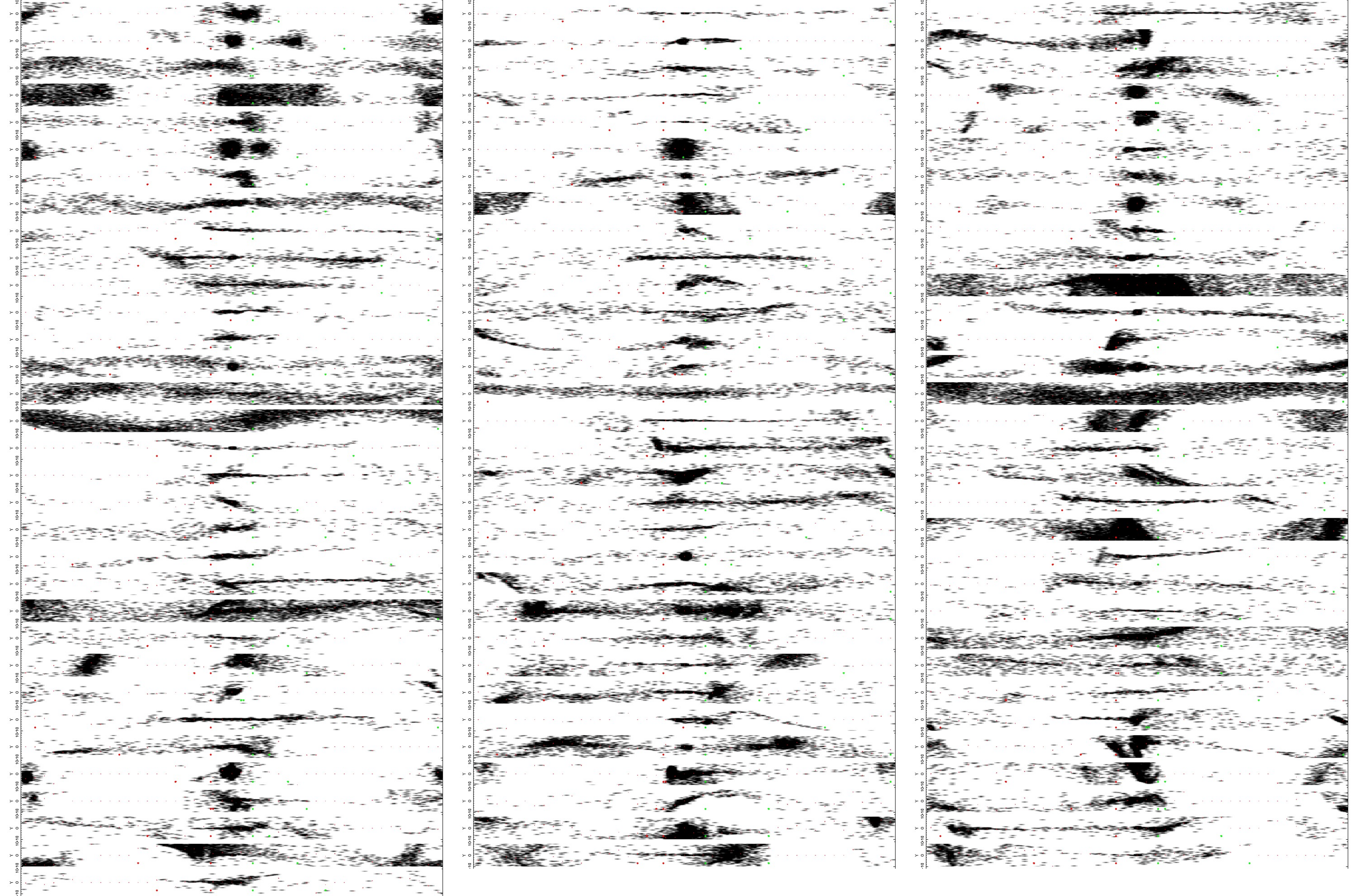}%{montage358_53.pdf}
\end{center}
\caption{Same as Figure~\ref{fig_in60} but allowing streams with RMS full widths of up to 5 degrees within 60 pc, which 
identifies 98 streams (last is not plotted).
}
\label{fig_in60_53}
\end{figure*}

\subsection{Running the Simulation}

The mixture of dark matter and star particles is evolved with Gadget2 \citep{Gadget2} and its updated version, Gadget3.
The n-body code has the added
Monte-Carlo routine to approximate the internal gravitational relaxation of the star clusters. 
The dark matter particles have
a softening of 200 pc. 

There typically are about 300,000 time steps in a simulation run from near redshift 3 to an age of 13.4 Gyr, which is
a simulation age of 13.7 units. 
A Milky Way-like halo dominates the final configuration, although there are many surrounding halos and a  halo with about 12\% of
the mass of the dominant halo is located about 1 Mpc from it.
The circular velocity in the main halo is 208 \kms\ at 10 kpc and peaks at 242 \kms\ around 35 kpc.

\section{Density Profiles of Streams}

\subsection{Time Evolution of a Stream}

The star clusters are inserted into sub-halos on nearly circular orbits. The dark matter density 
and the associated tidal field for an initial cluster orbit is about as strong as it will ever be 
with later merging into the main halo dropping leading to reduced  tides around most of the orbit \citep{Carlberg:18}.
The clusters therefore lose mass at a relatively high rate in their initial sub-halos. The tidal
stream spread out as a ring in the sub-halo, as shown in Figure~\ref{fig_sdm}. The online version of Figure~\ref{fig_sdm}
links to a movie showing the evolution of one cluster over the entire duration of the simulation.
The static figures and movie show that after a sub-halo falls  onto the main halo and merges,
the tidal stream that was within the sub-halo is jumbled and becomes a thick stream with a size 
characteristic of the orbit within the sub-halo. The differential velocities in the jumbled stream are 
roughly those of the orbit within the sub-halo, 5-10 \kms. Over half a Hubble time the velocities are sufficient 
to completely wrap the stars released at early times around the sky.
The cluster continues to lose stars to the tidal field which develops into a thin high density stream in the main halo 
as it settles down to a nearly stationary potential. The process is clearly visible in Figure~\ref{fig_sdm}, particularly 
in the movie.

Figure~\ref{fig_timeout} defines an instantaneous orbital plane using the velocity of the cluster (moving to the left in the figure) and projects the stars onto the sky using a cylindrical projection. 
The stars are color coded with the time since they left the cluster, defined as the last time they were within 100 pc of the
cluster center.  Stars that left within 0-3.01 Gyr  are black, 3.01-6.02 red, 6.02-9.03 green, and 9.03-12.04 (beginning of the simulation) blue. The stars in the thin stream are predominantly ones that left the cluster recently (black, red) 
and stars that are dispersed around the stream are predominantly ones that left earlier (green, blue). Essentially all clusters accreted onto the main halo have a similar time structure, with the numbers in the thin stream and the more dispersed component depending on the tidal history of the cluster and its mass. The lower planel of Figure~\ref{fig_timeout} shows
a cluster that has a much less prominent cocoon although virtually all the cocoon stars are old (blue) and the stream is dominated with young (black) stars.

The tidal history of three representative clusters is shown in Figure~\ref{fig_tides} with the cluster
of Figures~\ref{fig_sdm} and \ref{fig_timeout} shown in red. The tidal field for the red-line cluster is nearly constant
for the first 2 Gyr as it follows a nearly circular orbit in its sub-halo. When the sub-halo accretes onto the main
halo the sub-halo largely dissolves and the cluster begins a highly eccentricity orbit, with tides much weaker but the peak tides of comparable strength to those in the initial sub-halo. The cluster shown in green was quite close to the main halo and accreted at a very early stage on a nearly circular orbit. The tidal field declines slightly with time as the main halo slightly reduces its density.
The cluster shown in black has a similar history to the one shown in red, but with a much larger orbital radius, with apo-center beyond 150 kpc.  The values of the tides can be approximately related to orbital radius, $r$, with an isothermal
halo approximation $V_c^2/r^2$, with an approximate circular velocity of $230 \kms$, so the range shown corresponds
to a radial range of about 23 to 230 kpc.  The short-time spikes in the tidal fields are a result of sub-halo encounters with
the clusters.

An observational search for thicker streams will inevitably turn up unrelated field stars which also
formed within the same sub-halo and possibly even other clusters that were present.
Chemical tagging techniques may be able to identify the stars that are associated 
with individual star clusters which would allow
some reconstruction of the orbital velocity spread of the now dissolved dwarf 
and the relative time spent since the cluster's formation in the dwarf relative to
the time the cluster is free in the main halo. 

A numerical limitation of these particular simulations is that two body gravitational energy exchange between the 
star particles and the dark matter particles is expected to heat the tidal streams a few \kms\ over
a Hubble time in the main halo \citep{Carlberg:18}. 
The star-dark matter particle energy exchange scales as the inverse cube of the dark matter velocity dispersion, 
so the sub-halo streams
are expected to be heated significantly. If the dwarfs contain molecular clouds the heating 
would be approximately realistic, but it is 
not an accurately modeled process. Future simulations with larger numbers of particles will reduce the 
dark matter heating of tidal stream stars.

\subsection{Density along the streams}

The distribution on the sky of all the star particles between 8 and 40 kpc of the center 
of the main halo of the redshift 4.6 start simulation
are plotted with a Hammer-Aitoff (equal area) 
sky projection in 
Figure~\ref{fig_ah}. Alternate cutoff radii do not make much difference to the plot. 
The gray scale is proportional to the square root of the number of particles in each pixel of 0.002 units 
of the grid which ranges from $\pm 2\sqrt{2}$ in the horizontal direction and $\pm \sqrt{2}$ in the vertical direction,
hence a grid cell is approximately $0.13\degr$ on a side.
The gray scale saturates at 2.25 particles per pixel to help emphasize the lower density regions of the streams.  
Clusters in all stages of dissolution are present with
many clusters still within a sub-halo.
The overall distribution appears qualitatively similar
to the currently known streams on the sky \citep{Belokurov:06,PS1,GC:16,Shipp:18}. 
Two notable features are that most streams do not completely circle the sky, and, 
there is a substantial component of stars that are separated from the prominent thin streams.

Figure~\ref{fig_in60} shows the cluster by cluster orbital plane projections of all the particles 
from the 16 clusters within 60 kpc that
produce streams with an RMS full width along the stream of less than 2.4\degr. 
The width calculation does not subtract a local background as is often done with observational imaging data. 
There are more streams found as the maximum allowed RMS width is increased,   the numbers increasing to 29 at 3\degr\ full width, 
67 at 4\degr. and 98 at 5\degr.  
The first 97 streams of width 5\degr\ or less are shown
in Figure~\ref{fig_in60_53}.

\begin{figure*}
\includegraphics[angle=0,scale=2.1]{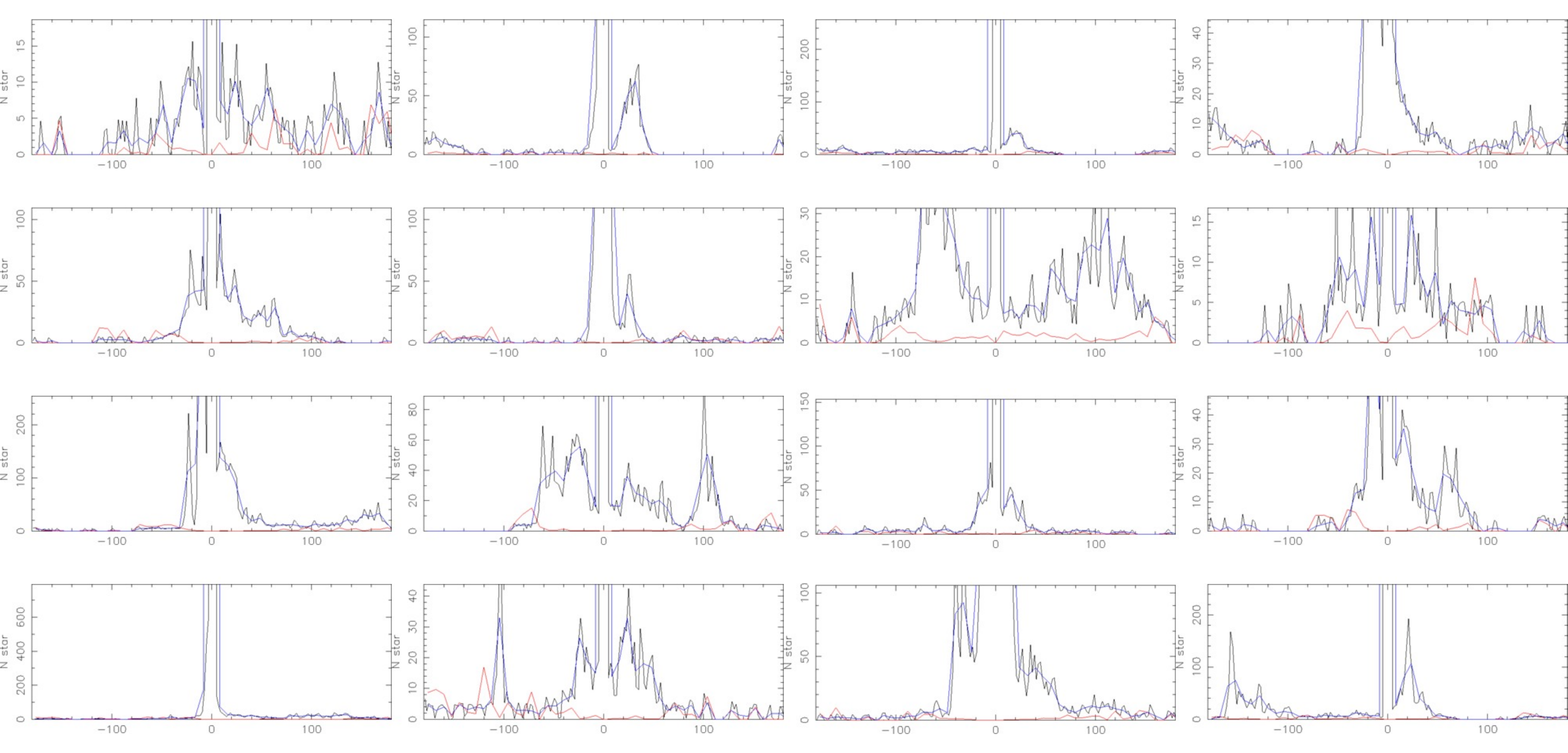}%{den358.pdf}
\caption{
The density along the 16 streams of Figure~\ref{fig_in60}.
The black lines are the density in 2\degr\ bins, the blue are the
density in 8\degr\ bins and the red is the estimated density variance in the 2\degr\ bins.
}
\label{fig_den358}
\end{figure*}

\begin{figure*}
\includegraphics[angle=0,scale=2.1]{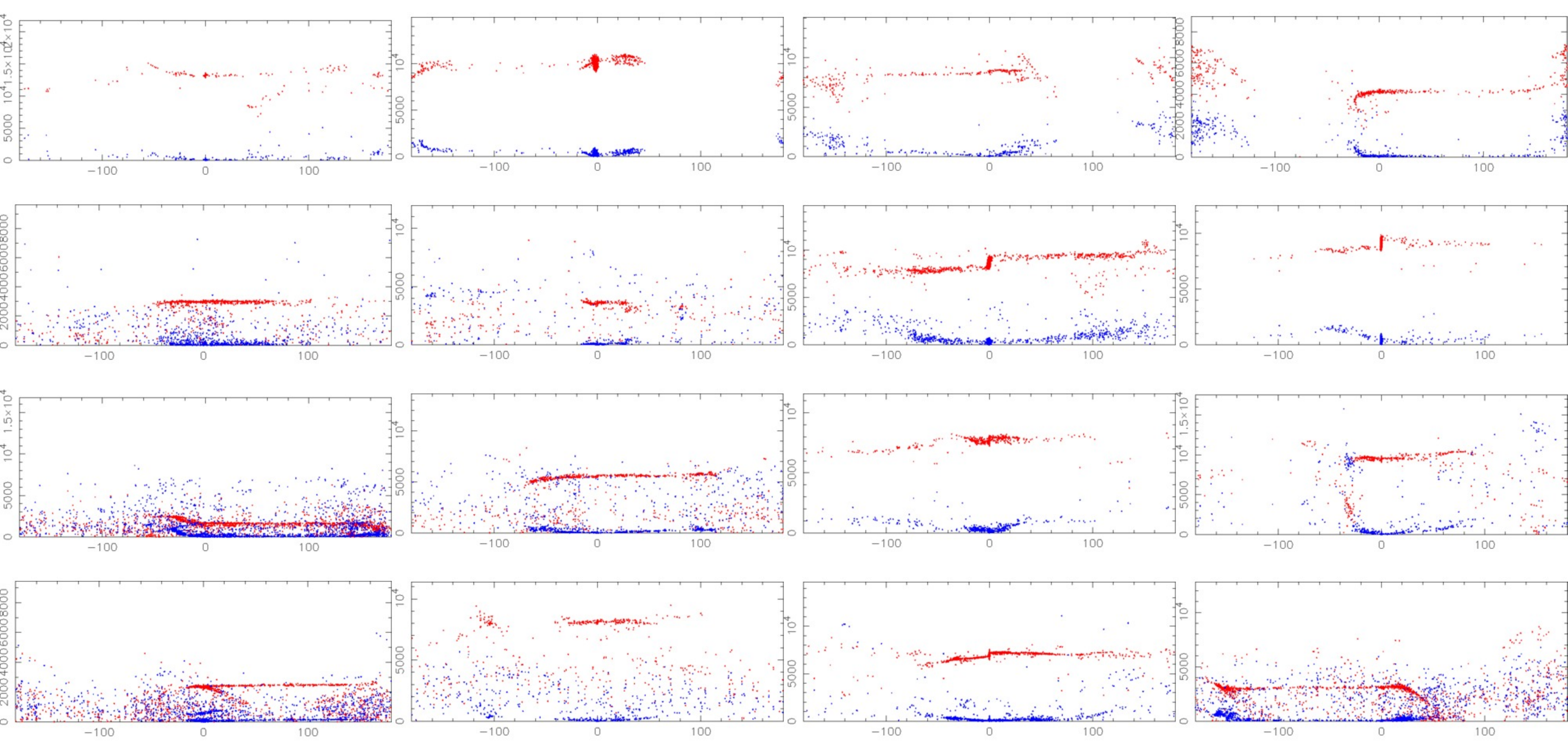}%{L358.pdf}
\caption{
The angular momentum aligned with the cluster (red) and the perpendicular component (blue) for the 
streams shown in Figure~\ref{fig_in60}.
The units are $\kms {\rm kpc}$.
}
\label{fig_L358}
\end{figure*}

The procedure to identify thin tidal streams in the simulations
in a manner comparable to observations first projects each stream onto its orbital plane on the sky.
The orbital plane is centered on the density maximum of the dominant dark matter halo. 
The velocity and position of the streams progenitor
star cluster defines an instantaneous angular momentum vector to which the nominal orbital plane is perpendicular. 
The stream will not lie exactly along the orbit that
the cluster defines even if the potential were spherical \citep{Sanders:14}, although for a completely
radial tidal field the offset will not be visible when viewed from the center of the galaxy.
The stream star particles are rotated into the new frame and projected onto
the sky in xy pixels $2\degr\times 0.3\degr$, smoothed with a 2D Gaussian with the same size and
shape as the pixels to give the streams of Figure~\ref{fig_in60}.

The particles within 1.2\degr\ of the high density centerline
are used to define the density of the streams as plotted in Figure~\ref{fig_den358}. The density around the entire 360\degr\ 
of the nominal orbital plane is plotted, whether or not a thin stream and a line of highest density can be identified.
It is readily evident that the stream densities are far from uniform with azimuth.
Many of the streams are on highly elliptical orbits which causes the density to decline and the stream to thin
at pericenter and pile up at apocenter. The density here is a mass density not a star count density which
will vary with the distance from the observer, the luminosity function of the stream and the density of background stars.
The black line of  Figure~\ref{fig_den358} is the density in 2\degr\ bins, the blue line is the density in 8\degr\ bins. 
The $\sqrt{n}$ variance in the counts is calculated from the 8\degr\ bins and then scaled by a factor of 2 to
the 2\degr\ bins and plotted as the red line. The region around the cluster itself is excluded from the calculation.

The stream search procedure used here requires that streams be at least 20\degr\ long and more than 12\degr\ from the progenitor cluster.
The angular length distribution of the streams of Figure~\ref{fig_in60}, as seen from the center of the galaxy, has a median of about 65\degr\
with the lower quartile at 46\degr\ and the top quartile at about 100\degr. The longest thin stream has an angular length of 150\degr.

\subsection{Angular momenta along the streams}

The instantaneous angular momentum of the star particles in the 16 thin streams of Figure~\ref{fig_in60}
is shown in Figure~\ref{fig_L358}. The  z direction is defined by the orbital angular momentum of the 
progenitor cluster.  The $L_z$ component of the particle angular momenta is shown in red and the two perpendicular 
components are summed in quadrature and plotted in blue. All stream particles are plotted, whether they
were found to be in the thin stream or not. The inner region of the dark halo halo is nearly spherical with the smallest to largest axes having
a ratio of about 0.97 at 30 kpc. Accordingly it is not surprising that the orbital momentum of the streams is approximately
constant along the length of the streams.

Figure~\ref{fig_L358} shows that there are star particles that have angular momenta 
well off the otherwise well defined line of the stream. There 
are two identifiable sources of these particles. First, there are the stars lost from the cluster before it 
merged into the main halo. Those are distributed around a ring in the sub-halo and therefore have an essentially
continuous variation of angular momentum when viewed from the center of the  main halo. The second
source of angular momenta scatter is the passages of dark
matter sub-halos through or near a stream. 
Given that the streams are orbiting at $\simeq 200\kms$ and that the numbers of  sub-halos only
become large enough to give a significant encounter rate 
below a sub-halo circular velocity of  $\simeq 5-10\kms$ the 
induced angular momenta changes will be about 2-5\% in a characteristic "sideways S" pattern.
Such changes appear to be present in some  streams and will be analyzed in a future paper.
Of course an encounter with a large sub-halo is also possible, although much less likely, which will
lead to much larger angular momenta changes.

\subsection{Transverse Density Profile of Streams}

\begin{figure}
\begin{center}
\includegraphics[angle=-90,scale=0.7,trim=90 30 40 30,clip=true]{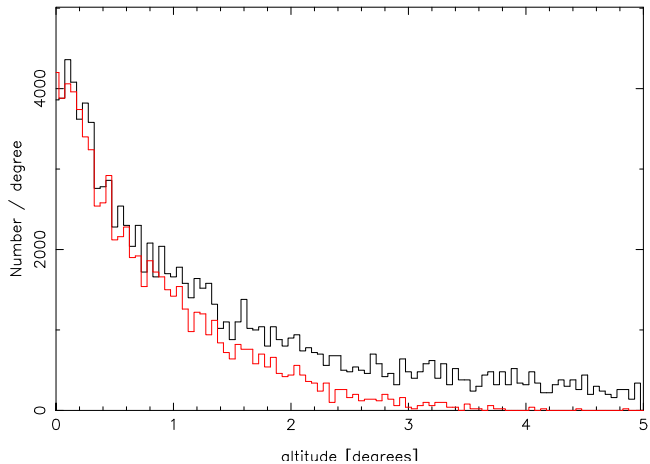}%{tden358.pdf}
\end{center}
\caption{The combined transverse density profile of all the streams in Figure~\ref{fig_in60}.
The red profile is for the particles identified as thin streams, the black is for all particles.
The red density profile shows an approximately Gaussian core of half width of about 0.4\degr\ and 
a much more extended, approximately exponential tail of half width of a degree or more.
}
\label{fig_tden358}
\end{figure}

Figure~\ref{fig_tden358} shows the density profile transverse to the 16 streams of Figure~\ref{fig_in60}.
The density relative to the centerline of particles identified as being in the thin streams is shown in red
and the density profile of all particles is shown in black.
The density profile shows an approximately Gaussian core of half width of about 0.4\degr. 
Figure~\ref{fig_tden358} also shows a much broader transverse density component, with a roughly exponential density profile with 
a half width of a degree or more, or a full width of 2-3\degr.

Figure~\ref{fig_tden358}  is a prediction that substantial numbers of particles should be 
distributed around the thin part of the streams. Methods which using imaging 
data alone and subtract a local background cannot find such broadly distributed streams.
However, there is some evidence that such wide components are being found
in streams with proper motion data \citep{PriceWhelan:18,Malhan:19}

\section{Effects of Reduced Low Mass Sub-Halo Numbers}

\begin{figure}
\begin{center}
\includegraphics[angle=0,scale=1.25,trim=200 200 160 200,clip=true]{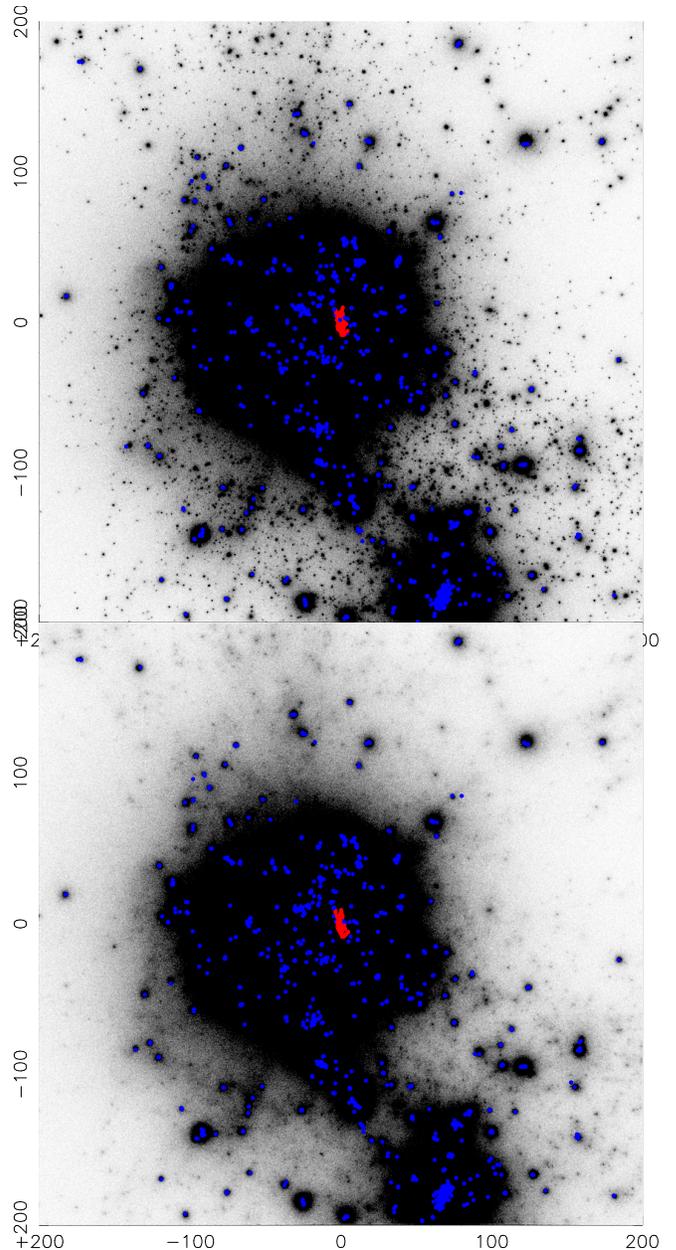}%{DM_4.pdf}
\end{center}
\caption{The dark matter distribution at 4 Gyr in boxes of 400 kpc for the simulation 
with normal low mass halos (top) and the inflated
low mass halos which dissolve in high density regions (bottom). 
The colored points show the location of the star clusters. 
The red points are a set of clusters in a single sub-dominant halo.
}
\label{fig_dm}
\end{figure}

The number of lower mass sub-halos orbiting in the Milky Way's halo is an interesting cosmological quantity.
Sub-halos in the mass range $\simeq 10^6-10^8\msun$ are sufficiently numerous  in the 30 kpc distance range and
sufficiently massive that there near stream passages will impart detectable velocity changes, which
lead to density variations.
The complication is the process of sub-halo merging into the main halo also has substantial
potential fluctuations over a wide range of scales that it too is expected
to create density variations in the streams. A straightforward test of the effect of the lower mass halos is to rerun the same
simulation, but without the lower mass halos present and compare the density fluctuations in the with and without simulations.

A model with suppressed low mass halos must have the same total mass and distribution of mass to ensure that 
two simulations have largely the same overall assembly history. After a number of trials, the best
sub-halo suppression technique is to inflate the radius of
the low mass halos, a factor of 4 expansion producing relatively weakly bound sub-halos without excessive overlap with other halos.
The velocity dispersion of the halos is reduced proportionally, a factor of 2, 
so that the low mass sub-halos start in equilibrium if there were no outside forces.
The mean density of the inflated
low mass halos is therefore reduced by a factor of 64, which means that they
quickly disperse once they enter any higher mass virialized halo.
The dark matter distribution is shown for the two models in Figure~\ref{fig_dm}.

\begin{figure}
\includegraphics[angle=-90,scale=0.8,trim=50 30 10 30,clip=true]{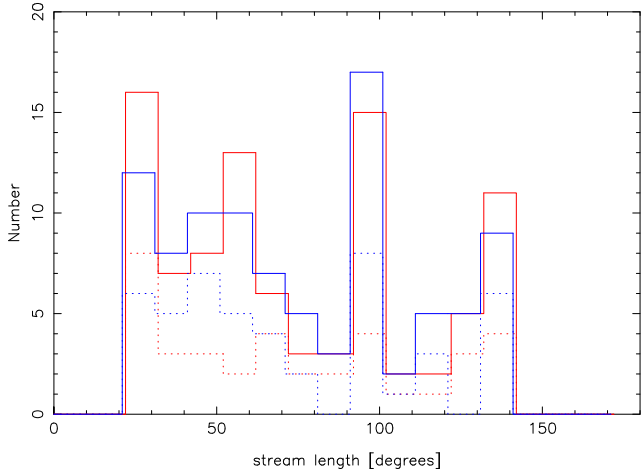}%{Snlen.pdf}
\caption{The distribution of the  lengths of the thin streams in the two simulations to a distance of 100 kpc.
The total length is from left to right of the cluster.
 Red is normal sub-halos, blue is suppressed
low mass sub-halos. The dashed line removes streams that come within 15 kpc of the center.
}
\label{fig_Snlen}
\end{figure}

The clusters for the study of the response of streams to different sub-halo populations 
have somewhat different parameters than the run discussed in the previous section to try to boost the number
of streams and to lower the particle noise in the stream measurements.
The star masses are increased to $10\msun$ from $5\msun$ allowing twice as many clusters to be creates, 1433 
instead of 710. And, the cluster internal heating model is an older version which causes significantly more mass
loss from clusters above $10^5\msun$ which has the benefit of boosting the numbers
of star particles in the streams. The model is started from redshift 3.2. 

Figure~\ref{fig_Snlen} shows the distribution of stream lengths in the with and without lower mass sub-halo simulations, for streams whose most distant
star particles are at 150 kpc
of the galactic center. The dotted lines show the distribution excluding streams that come within 15 kpc of the 
galactic center, with the solid lines showing all streams. The length distributions of the streams
in the two simulations are essentially identical and give confidence that the basic large scale properties 
of the streams in the with and without lower mass sub-halo simulations are largely identical  so
that comparing the smaller scale properties is a useful test.

\begin{figure}
\begin{center}
\includegraphics[angle=-90,scale=0.8,trim=50 30 10 30,clip=true]{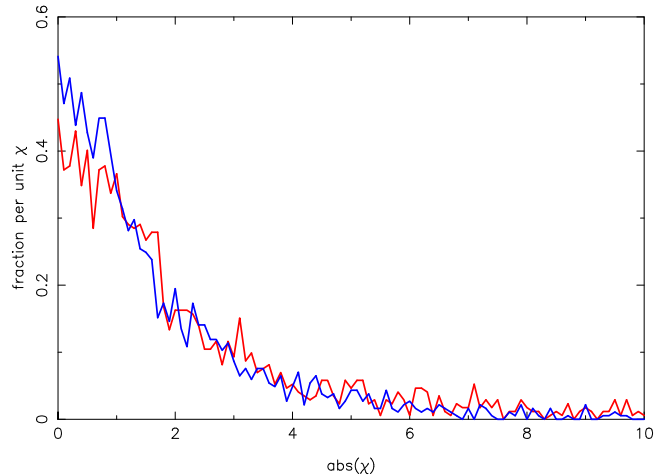}%{Schi.pdf}
\end{center}
\caption{The distribution of normalized density variations in a thin stream when measured in 4\degr\ bins. 
The standard sub-halo population simulation is  the red line and the simulation with suppressed  
sub-halos below $4\times 10^8\msun$, 
the  blue line.
}
\label{fig_schi}
\end{figure}

\begin{figure}
\begin{center}
\includegraphics[angle=-90,scale=0.8,trim=50 30 10 30,clip=true]{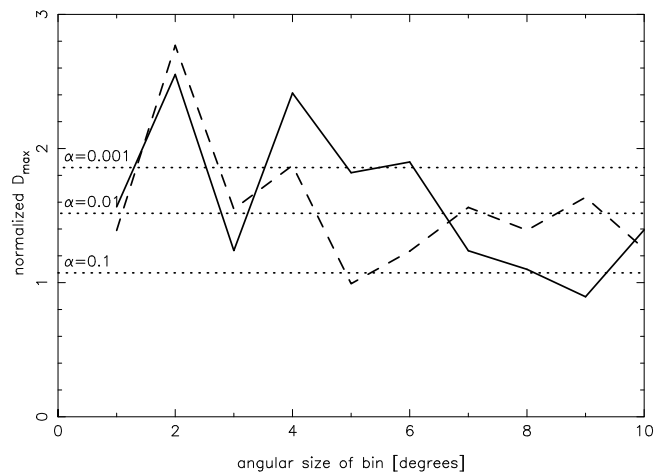}%{KS.pdf}
\end{center}
\caption{The Kolmogorov-Smirnov test statistic for the with and without sub-halo 
simulations as a function of the angular size of the bins in which
the density variation is measured. The vertical scale is statistical probability that the result occurs by chance.
The test is done at the end of the two simulations and at a time 
0.5 Gyr earlier (dashed line), to give an indication of  the time variation of the result.
}
\label{fig_ks}
\end{figure}

\subsection{Effect of Sub-halos on Stream Density Variations}

Tidal streams in an effectively static potential  have substantial structure in the plane of the orbit, especially  near 
the progenitor cluster \citep{Kupper:08,Kupper:12} as a result 
of variations in mass loss rate around the orbit. However when viewed from 
within the plane of the orbit the tidal mass loss features increasingly overlap
with distance and the unperturbed stream is relatively smooth \citep{Carlberg:15,WB:19}.
Even for a Milky Way-like galaxy whose accretion history over the last half Hubble time 
has been fairly quiet with no major mergers, the potential varies substantially which needs to be taken into account
to understand tidal stream structure.
The simulations here are used to measure the
small scale stream density variations at the current epoch of a realistic Milky Way assembly model with 
and without lower mass sub-halos.

A simple non-parametric approach to measure the fraction of streams with density variations is 
to compute the $\chi^2$ per degree of freedom, $\nu$,
for the thin stream densities. 
The varying local mean density is estimated using a fourth order fit, $m(\phi)$, to the density which also gives
an estimate of the local statistical variance in the density. Consequently, 
 $\chi(\phi) = (d(\phi)-m(\phi))/\sqrt{m(\phi)}$ and these values are quadrature summed over 
each identified streams. 
The outcome is that about 70\% of both the 
with and without sub-halo simulations  have $\chi^2/\nu>2$, and, about 50\% of the streams have $\chi^2/\nu>5$ with 
differences of only 4\% between the fractions in the two simulations.
No formal confidence level is 
associated with these values because the distribution of densities has a non-Gaussian tail.
Therefore,
the presence of stream density variations has no significant dependence on the presence, or not, 
of sub-halos below $4\times 10^8 \msun$ in the simulation.

The combined total length of thin streams over which density measurements can be made is about 7400\degr\
in both two simulations. 
Figure~\ref{fig_schi} shows the distribution of the individual $\chi(\phi)$ values when
measured in 4\degr\ bins of azimuth along the streams. 
The simulation with sub-halos has an RMS spread of the $\chi$ values that is 22\% larger than the simulation without low mass sub-halos.

The cumulative distributions of the $\chi(\phi)$ values for the
simulation with and without sub-halos are used to compute the Kolmogorov-Smirnov statistic, $D_{max}$, 
which is simply the greatest difference
between the two cumulative probability distributions  \citep{NRC:92}. Counts-in-cells is a useful approach to take 
because the local values are helpful in identifying the location in  the stream of strong density variations,  
which may be where sub-halos may have passed. The counts are directly related to both the two-point correlation function and 
the power spectrum, with the variance of the counts in cells is the sum of the Shot noise plus 
the two-point correlation function integrated
over the size of the bin.  The two-point correlation function is the Fourier transform of the power spectrum.

Figure~\ref{fig_ks} shows the KS  statistic, $D_{max}$, normalized with the sample size factor 
$\sqrt{n_1n_2/(n_1+n_2)}$ for the two samples. The normalization
allows different angular bin sizes, hence sample sizes, to be compared on a common plot. At 2 and 4 degrees the simulation 
with sub-halos shows a highly significant excess in density fluctuations 
over a simulation without sub-halos. At angles out to 6 degrees the statistical significance 
is reduced but significant differences between the two distributions are present.
These results are a useful basis for further effort to statistically identify a sub-halo population in the Milky Way. 
These angular scales are about
what one expects sub-halos with scale radii in the range of 0.5-1 kpc to induce in a stream observed at 
a typical distance of about 20 kpc. 
Figure~\ref{fig_ks} also shows the same comparison done 0.5 Gyr earlier to provide a rough estimate 
of the time variation of the differences.
The larger fluctuations in the 2-4\degr\ range appear to be a consistent feature \citep{BB:19}. It is important
to note that 
Figure~\ref{fig_schi} shows that even with nearly 8000\degr\ of stream measurements in hand
the difference in the distribution of the $\chi$ values is not dramatic even when a well matched pair
of models is available.

\section{Discussion and Conclusions}

This paper explores the density structure of tidal streams from globular clusters that form in sub-galactic halos and then are
incorporated into the galactic halo. The longer lived clusters lose mass in their sub-galactic halo which is spread out over the orbit of 
the cluster. Once the sub-halo falls into the main halo those stars are spread out in a thick stream around the thin stream which forms.
The density 
profile transverse to the streams, Figure~\ref{fig_tden358}, shows a Gaussian core of full width of about 0.8\degr\ and then a roughly
exponential tail of stars in a wider stream with a full width half maximum of about two degrees. The existence
of a significant wider stream component in tidal streams is expected on the basis that most clusters 
formed in sub-galactic halos which later merged into the main halo of the galaxy. 

To examine the role of low mass halos in causing density variations in the thin streams
matched simulations with and without dark matter sub-halos below $4\times 10^8 \msun$ suppressed are compared. 
The fraction of streams that show large density fluctuations 
has effectively no dependence on whether lower mass
sub-halos are present or not.  However, binning the density along the stream in different angular size
bins finds that there is a highly significant excess in the density fluctuations on scales of 2-6\degr\ relative
to a well matched model without low mass sub-halos.

The underlying purpose of this paper is first to generate realistic simulations of globular cluster tidal streams in a galaxy
with an assembly history roughly like the Milky Way, in which the halo has not had any major mergers for the last half Hubble time. 
The simulations are useful tests of methods to identify dark matter sub-halos in the Milky Way. Even though individual density features
are not reliable tests for the presence of the lower mass sub-halos,
a large set of density measurements for streams may be. A future analysis will examine
the velocity variations in these streams and their dependence on the sub-halos.

\acknowledgements

Comments from a referee led to substantial improvements in this paper. I thank Khyati Malhan for discussions.
This research was supported by  NSERC of Canada. Computations were performed on the niagara supercomputer at the SciNet HPC Consortium. 
SciNet is funded by: the Canada Foundation for Innovation; the Government of Ontario; 
Ontario Research Fund - Research Excellence; and the University of Toronto.

\end{document}